\documentclass[12pt]{article}
\usepackage[textwidth=6.25in]{geometry}

\usepackage{enumerate}
\usepackage{amsmath} 
\usepackage{times}
\usepackage{amsmath,amsthm,amssymb}
\usepackage{fancyhdr}
\usepackage{moreverb}
\usepackage{graphicx}
\usepackage{amssymb}
\usepackage{url}
\usepackage{multirow} 
\usepackage[boxed]{algorithm}
\usepackage{algpseudocode}
\usepackage{multirow,bigdelim}
\usepackage{geometry}
\usepackage{fix-cm}
 \usepackage{setspace}
\usepackage{natbib}
\usepackage{color}
\usepackage{bbold}
\usepackage{graphicx}
\usepackage{caption}
\usepackage{subcaption}
\usepackage{sectsty}
\usepackage{titlesec}
\usepackage{pifont}%
\titleformat{\section}[hang]{\centering \bfseries\large}{\thesection.}{0.4em}{}

\usepackage{tikz}
\usetikzlibrary{arrows}

\newcommand{\R}{\mathbb{R}}

\newcommand{\supp}{\text{supp}}

\renewcommand{\P}{\mathbb{P}}
\newcommand{\E}{\mathbb{E}}

\newtheorem{theorem}{Theorem}[section]

\newtheorem{corollary}[theorem]{Corollary}

\DeclareMathOperator*{\argmax}{arg\,max}
\DeclareMathOperator*{\argmin}{arg\,min}

\theoremstyle{definition}
\newtheorem{example}[theorem]{Example}

\theoremstyle{definition}

\theoremstyle{assumption}
\newtheorem{assumption}[theorem]{Assumption}

\theoremstyle{class}
\newtheorem{class}[theorem]{Definition}

\theoremstyle{collection}

\def\one{{\bf 1}}

\def\bSigma{{\bf \Sigma}}

\def\v{{\bf v}}

\def\w{{\bf w}}

\def\X{{\bf X}}

\def\bBeta{\boldsymbol{\beta}}

\def\F{{\cal F}}
\def\G{{\cal G}}
\def\H{{\cal H}}
\def\P{{\mathbb P}}
\def\E{{\mathbb E}}
\def\Var{{\rm Var}\,}
\def\Cov{{\rm Cov}\,}
\def\Corr{{\rm Corr}\,}

\def\|{\, | \,}

\def\diag{\text{diag}}

\def\diag{\text{diag}}

\def\logit{{\rm logit}}
\def\probit{{\rm probit}}

\allowdisplaybreaks

\title{Combining Information from Multiple Forecasters: Inefficiency of Central Tendency}
\author{Ville A. Satop\"a\"a\thanks{Ville A. Satop\"a\"a is a Statistician, Department of Technology and Operations Management, INSEAD, Boulevard de Constance, 77305 Fontainebleau CEDEX, France (e-mail: ville.satopaa@insead.edu). The author wishes to acknowledge and thank the support of the MacArthur Foundation's Opening Governance Network.}
}
\date{}

\begin{document}
\maketitle

\begin{abstract}
\singlespace

Even though the forecasting literature agrees that aggregating multiple predictions of some future outcome typically outperforms the individual predictions, there is no general consensus about the right way to do this. Most common aggregators are means, defined loosely as aggregators that always remain between the smallest and largest predictions. Examples include the arithmetic mean, trimmed means, median, mid-range, and many other measures of central tendency. If the forecasters use different information, the aggregator ideally combines their  information into a consensus without losing or distorting any of it. An aggregator that achieves this is considered efficient. Unfortunately, our results show that if the forecasters use their information accurately, 
an aggregator that always remains strictly between the smallest and largest predictions is never efficient in practice. A similar result holds even if the ideal predictions are distorted with random error that is centered at zero. 
 If these noisy predictions are aggregated with a similar notion of centrality, then, under some mild conditions, the aggregator is asymptotically inefficient.
%
\end{abstract}


\section{INTRODUCTION} \label{introduction}
Forecasters often use different information in their predictions. For instance, one analyst may live in Russia while another lives in the United States. If they both follow local news, they are likely to use very different information in their predictions about global political events.   Ideally their predictions would represent an accurate assessment of their information. 
In the economics and statistics literature such ideal predictions are called rational, honest, or \textit{calibrated} (see, e.g., \citealt{keane1990testing, dawid1995coherent, ottaviani2006strategy, Ranjan08, jolliffe2012forecast, satopaamodeling}). Note that even though this paper treats the forecasters as human experts, the discussion is equally valid for other types of forecasters, such as statistical models.

The predictions are then combined with a statistical aggregator that is nothing but a function from the predictions to the consensus. This is typically done for the sake of decision making and improved accuracy. In fact, after summarizing a wide range of empirical and simulated evidence in the forecasting literature, \cite{clemen} states that ``the results have been virtually unanimous: combining multiple predictions leads to increased prediction accuracy."  In general, the best that an aggregator can do is to use all information that the forecasters provide through their predictions. If the aggregator does this without losing or distorting any of the forecasters' information, it is considered \textit{efficient}.
There are, however, many different ways to aggregate predictions, and the choice of the aggregator can largely determine how well the consensus performs out-of-sample \citep{satopaamodeling2}. 
Often the first idea is to use a \textit{mean}, defined here as any aggregator that is bound to remain between the smallest and largest predictions.
This definition is due to \cite{cauchy1821cours} and is probably the least restrictive definition of a mean. Given that in colloquial language ``mean'' and ``arithmetic mean''  (i.e.,  the sum of values divided by the number of values) are often treated as synonyms, it is important to emphasize that our scope is not restricted to the arithmetic mean
but instead captures all means including the weighted arithmetic mean, median, mode, mid-hinge, mid-range, trimmed means, geometric mean, harmonic mean, and many others. 

Today the practice of combining predictions with a mean appears almost ubiquitous. For instance, central banks use the arithmetic mean or median to combine macroeconomic predictions of inflation and real output growth \citep{trabelsi2016central}; the Survey of Professional Forecasters by the Federal Reserve Bank of Philadelphia reports the arithmetic mean and median predictions of real gross domestic product (GDP), unemployment rate, and other macroeconomic indicators; Reuters.com, Yahoo! Finance, Fortune, and other business sites show the arithmetic mean of analysts' predictions of financial indicators such as future sales and earnings; and many others. In fact, it is rather challenging to find a real example that does not use a mean to aggregate predictions.
%




Unfortunately, if the goal is to find or provide the public with the most informed consensus prediction, the common practice is falling short. This is because means are not necessarily efficient. To illustrate with a very simple example, 
suppose a patient has a magnetic resonance imaging (MRI) and blood test taken. Two well-calibrated doctors then independently analyze the results that fortunately show no indication of poor health. The first doctor only looks at the MRI results and predicts that the patient has a probability $0.9$ of being healthy. The second doctor, on the other hand, only looks at the blood test results but also predicts a probability $0.9$ of the patient being healthy. Now the patient has two predictions of $0.9$, each based on very different information. 
How should these be combined? Surely, if one looked at both the MRI and blood test results, one would be even more confident of the patient's good health and hence predict a probability somewhat greater than $0.9$.  In this simple scenario, however, all means aggregate to $0.9$. They fail to account for the doctors' differing information sets and hence are inefficient. 

Even though this example may appear somewhat contrived, it does suggest an interesting question: under what conditions are means efficient aggregators of point predictions that are based on different information? Strikingly, our main result shows that  the answer is ``almost never.''


The current paper develops this result with the following structure. First, Section \ref{propertiesS} introduces the framework of analysis, gives a brief literature review, and derives the necessary and sufficient conditions for efficiency (Theorem \ref{optimal}). The discussion considers a \textit{weak form of efficiency} in a sense that an aggregator is efficient if and only if there exists some probability space under which it uses all the information in the predictions. Failing to meet this criterion then shows a \textit{strong form of inefficiency}; that is, an aggregator is inefficient if there is no probability space under which it uses all the information in the predictions. 

Section \ref{MeasuresOfCentralTendency}  shows that this type of inefficiency applies to means of calibrated predictions about continuous (GDP, sales, stock prices), binary (rain or no rain, candidate wins or loses), discrete (product demand, goals per season), and all other numerical types of univariate outcomes. The result is developed in two parts:

\begin{enumerate}[a)]


\item Section \ref{contractionSection} analyzes arithmetic means.
An arithmetic mean is non-trivial and attempts to aggregate information if it assigns positive (fixed) weight to at least two different predictions.
Theorem \ref{contraction} shows that a non-trivial arithmetic mean violates all the efficiency properties  in Theorem \ref{optimal} and is therefore inefficient. 
 More specifically, it distorts the forecasters' information and is under-confident.


Sometimes the arithmetic mean is believed to be inefficient because it represents the forecasters' shared information too many times \citep{kim2001inefficiency,palley2015eliciting}. This, however,  is not an entirely accurate explanation. 
We show that a linear aggregator faces a trade-off between \textit{under}-representing private information (known only to one forecaster) and \textit{over}-representing the shared information (known to all forecasters). The efficient strategy balances the trade-off by intentionally over-representing the shared information in order to under-represent each forecasters' private information less. 
The arithmetic mean always represents shared information optimally but unfortunately under-represents any private information. This suggests that the arithmetic mean can work well  when the forecasters rely on very similar information; otherwise a relatively large share of information is under-represented, leading to an under-confident aggregate.

\item Section \ref{RandomSection} analyzes the general class of means. 
This class is not entirely inefficient but if the definition is tightened slightly such that a mean must  
stay strictly between the smallest and largest predictions whenever at least two predictions disagree, inefficiency becomes a class property. 
More specifically, Theorem \ref{centralTendency} shows that these aggregators, that we call \textit{strict means}, can be efficient only if the forecasters' information sets are  uncountably large. Given that in our finite physical universe forecasters use finite information sets, strict means and hence essentially all common aggregators are not efficient in practice.

\end{enumerate}


Section \ref{noisy} applies these results in an asymptotic analysis with forecasters who make mistakes in interpreting their information and hence report noisy versions of their calibrated predictions. In other words, each prediction is a draw from a distribution with the center at that forecaster's calibrated prediction. If the predictions are aggregated with a similar notion of centrality, Corollary \ref{asympEff} shows, under some mild conditions, that the aggregator is asymptotically inefficient.
Section \ref{conclusion} concludes by discussing our results in the broader context of forecast modeling, intuitively explaining the reason behind the inefficiency of central tendency, and describing some of our future work in forecast aggregation. 
\section{PARTIAL INFORMATION FRAMEWORK} \label{propertiesS}

\subsection{Predictions and Outcome} \label{outcomes}
The partial information framework is a micro-level model of different predictions \citep{satopaamodeling2}. More specifically, suppose there are $N \in \mathbb{N}$ forecasters making point predictions $X_1, \dots, X_N$ about some future outcome $Y$. The predictions and outcome are treated as random variables defined under a common probability space $(\Omega, \F , \P)$. In general, assuming the existence of a probability space cannot be avoided in any probabilistic treatment of forecast aggregation. In fact, according to \cite{lindley} this is essential and ``any other approach has an element of adhockery about it [...].'' 
Under the given probability space, define $\E(Y) := \mu_0 < \infty$ and $\Var(Y) := \delta_0 < \infty$. This leads to the following assumption about the outcome:
\begin{assumption}
The outcome $Y$ is defined under $(\Omega, \F , \P)$ and $Y \in \mathcal{L}^2$, where $\mathcal{L}^p  := \{ Z \in \F : \E[|Z|^p] < \infty\}$\label{Assump1}
\end{assumption}


\noindent

Each forecaster predicts $Y$ based on some partial information. In mathematics information is often formalized with $\sigma$-fields  \citep{herves2013information}. This way the trivial $\sigma$-field $\F_0 := \{\emptyset, \Omega\}$ represents no information and the principal $\sigma$-field $\F$ represents all possible information that can be known about $Y$. The $j$th forecaster, however, knows some partial information set  $\F_j \subseteq \F$. No assumption as to how the forecaster acquired this information is needed. For instance, it could  have stemmed from books, news articles, personal recollections, and so on (see \citealt{satopaamodeling} for futher discussion). These details are not relevant for aggregation and hence can be abstracted away.


The $j$th forecaster first forms a conditional distribution $\P(Y|\F_j)$.  How the forecaster then converts this distribution into a point prediction $X_j$ depends on the
 loss function $L(X,Y)$ used for evaluating prediction accuracy.
  The partial information framework assumes that the forecasters aim to minimize the expected \textit{Bregman loss}, defined as follows: if $\phi: \R \to \R$ is a strictly convex, differentiable function, then the Bregman loss function is $L(x,y) = \phi(x)-\phi(y) - (x-y)\phi'(y)$. For instance, $\phi(x) = x^2$ gives the squared error loss $L(x,y) = (x-y)^2$. In general, if $\phi$ is continuously differentiable, then Bregman loss functions are precisely those functions whose expected value is minimized by $\E(Y|\F_j)$ among all $\F_j$-measurable random variables, i.e.,  $\argmin_{X \in \F_j} \E[L(X,Y)] = \E(Y|\F_j)$. 
This leads to the following assumption about the predictions:
\begin{assumption}
The prediction $X_j = \E(Y | \F_j)$ a.s. for some $\F_j \subseteq \F$. \label{assumption2}
\end{assumption}



In practice the forecasters may not use all the information they know about $Y$. For instance, they may not be able to  reconcile all the available information or they may conceal some of it for various reasons. Either way, the aggregator can access and operate only on the information that the predictions reveal to it. Therefore, for the sake of aggregation, any unused information may as well not exist, and it is in no way restrictive to set $\F_j = \sigma(X_j)$ and let $X_j = \E(Y | X_j)$ a.s. 

This condition, typically known as calibration, is a common assumption in theoretical analysis of forecasting (see, e.g., \citealt{keane1990testing, dawid1995coherent, Ranjan08, jolliffe2012forecast, satopaamodeling}). Calibration is equivalent to Assumption \ref{assumption2}; that is,  a prediction $X_j$ is calibrated if and only if there exists some information set $\F_j \subseteq \F$ such that $X = \E(Y|\F_j)$ a.s. \citep{satopaamodeling}. Given that our definition of calibration depends on the probability measure $\P$, it may seem  different from the widely studied notion of empirical calibration (\citealt{dawid1982well, foster1998asymptotic}; and many others). 
However, as was pointed out by \citet{dawid1995coherent}, these two definitions of calibration can be expressed in formally identical terms by letting $\P$ represent the limiting joint distribution of the outcome-prediction pairs. 


 To illustrate calibration, consider a probability prediction $X_j \in [0,1]$. If the outcome happens $100p\%$ of the time among all those times that $X_j = p$, then the prediction is calibrated. This is precisely how probabilities are generally understood.  Most forecasters with a basic understanding of probabilities are likely to target good calibration even without being instructed to do so. 
In fact, previous studies have found many calibrated subpopulations of experts, such as meteorologists \citep{murphy1977reliability}, experienced tournament bridge players \citep{keren1987facing}, bookmakers \citep{dowie1976efficiency}, and  economic forecasters \citep{keane1990testing}. 

On the more theoretical side, \cite{dawid1982well} show that an internally consistent Bayesian can expect to become calibrated under feedback. \cite{foster1998asymptotic} show that the forecaster needs to know almost nothing about the outcome in order to be calibrated.  If one is nonetheless left with uncalibrated predictions, it may be possible to calibrate them either empirically or with an appropriate model \citep{dawid1995coherent, satopaamodeling2}. Furthermore, the calibration of human forecasters can be improved through team collaboration, training, tracking \citep{mellers2014psychological}, timely and accurate feedback \citep{murphy1984impacts}, representative sampling of target events \citep{gigerenzer1991probabilistic, juslin1993explanation}, breaking the task into smaller more manageable sub-parts \citep{tetlock2016superforecasting}, or by evaluating the forecasters under a 
Bregman loss function \citep{banerjee2005optimality}.

\subsection{Efficient Aggregators}\label{InfoCollection}
Throughout this paper aggregators are denoted with different variants of the script symbol $\mathcal{X}$. An aggregator is simply any prediction that is measurable with respect to $\F'' := \sigma(X_1, \dots, X_N)$, namely the $\sigma$-field of all information that can be inferred from the reported predictions
\citep{Ranjan08}.
If the aggregator uses all of this information, it is called efficient. The following definition makes this precise. 





\begin{class}[Efficiency]\label{infoAgg}
An aggregator $\mathcal{X}$ is efficient if there exists a probability space $(\Omega, \F, \P)$ under which $\mathcal{X} = \E(Y | X_1, \dots, X_N)$ a.s. or, equivalently, $\mathcal{X} = \E(Y | \F'')$ a.s. with $\F'' = \sigma(X_1, \dots, X_N)$. Denote all efficient aggregators with $\mathcal{X}''$. 
\end{class}
\noindent

In regression analysis $\mathcal{X}''$ is known as the \textit{regression function} and is often considered the \textit{best predictor} of $Y$ (see, e.g., \citealt[Theorem 6.3.1]{christensen2011plane}) because it minimizes the expected Bregman loss \citep{satopaamodeling2}. 
Unfortunately, Definition \ref{infoAgg} is rather abstract and does not render itself well to analysis. In response, the next theorem fully characterizes efficient aggregators in terms of two necessary and sufficient properties. 
These properties form a ``toolbox'' for 
showing that some aggregator is not efficient.
In particular, violating any of the properties shows a strong form of inefficiency: there is no probability space under which the aggregator is efficient. The proofs of this and all other theorems are deferred to the Appendix.







\begin{theorem} \label{optimal}
An aggregator $\mathcal{X}$ of calibrated predictions $X_1, \dots, X_N$ is efficient if and only if it has the following two properties.
\begin{enumerate}[(i)] \label{properties}
\item $\mathcal{X}$ is \textit{calibrated}: $\E(Y|\mathcal{X}) = \mathcal{X}$ a.s. \label{calibrationEfficient}
\item $\mathcal{X}$ is \textit{extremizing}: $\Var\left[ \E(Y | X_j,  j \in \v)\right] \leq \Var(\mathcal{X})$ for all  subsets $\v \subseteq \{1, \dots, N\}$. \label{expandEfficient}
\end{enumerate}
\end{theorem}
Item (\ref{calibrationEfficient}) does not require calibrated predictions. In fact, an efficient aggregator, like any conditional expectation, is always calibrated. Given that $\mathcal{X}''$ is calibrated, it is also marginally consistent: $\E(\mathcal{X}'') = \E(Y) = \mu_0$. To see this, consider a calibrated prediction $X$. Then $\E(X) = \E[\E(Y|X)] = \E(Y) = \mu_0$. Consequently, all calibrated predictions (individual or aggregate) are marginally consistent. The converse, however, is not true. For instance, Theorem \ref{contraction} shows that any non-trivial arithmetic mean is marginally consistent but not calibrated. This is an important observation because it provides a technique for proving lack of calibration (and hence inefficiency) via marginal inconsistency -- a task that can be much easier than proving lack of calibration directly. 




Item (\ref{expandEfficient}), on the other hand, requires calibrated predictions. 
To interpret this property, suppose the forecasters' information sets form an increasing sequence
$\{\emptyset, \Omega\} = \F_0  \subseteq \F_1 \subseteq \dots \subseteq \F_N \subseteq \F$.  Note that $\E(Y|\F_0) = \mu_0$ and $\E(Y|\F) = Y$. \citet{satopaamodeling2} show that the variances of these predictions follow the same order as their nested information sets, that is, $0 = \Var(X_0) \leq \Var(X_1) \leq \dots \leq \Var(X_N) \leq \Var(Y) = \delta_0$. Thus the informativeness of a calibrated prediction increases in its variance. Naturally, if an aggregator is efficient, it uses all of the forecasters' information and hence at least as much information as any subgroup of these individuals. Therefore $\mathcal{X}''$ is necessarily at least as variable as every individual prediction or, more generally, the efficient aggregate of any subset of them. 
This is precisely what item  (\ref{expandEfficient}) describes. 

\subsection{Literature Review}\label{Lit Review}

One of the earliest inefficiency results is by Nobel laureate C. W. J. Granger who emphasized the importance of forecasters' information in both theory and interpretation of predictions \citep{granger1989invited}. He shows that the equally weighted arithmetic mean is an inefficient aggregator of calibrated predictions if each forecaster's information consists in equal parts of shared and private information. Later on, \cite{kim2001inefficiency} reached the same conclusion. 



Real-world forecasters, however, can have different amounts of information, and not all information is either completely shared or private. For instance, two forecasters may share information that a third forecaster does not know and so on. 
This way the forecasters' information can be seen to form a complex structure of partially overlapping sets. 
By operating under the partial information framework, it is possible to analyze aggregation without constraining the information structure. Some previous authors have taken this approach to examine calibrated forecasting of binary outcomes. In particular, \cite{dawid1995coherent} show that the weighted arithmetic mean of two calibrated predictions of a binary outcome is always inefficient as long as one forecaster does not know everything that the other forecaster knows.
This result was later on extended to $N \geq 2$
calibrated predictions of a binary outcome \citep{hora2004probability, Ranjan08, gneiting2013combining}. 

  \cite{parunak2013characterizing} attempt to generalize beyond the arithmetic mean. They suppose that all forecasters know equally much and that not all information about a binary outcome can be known. They then consider an omniscient prediction that optimally uses all information that could possibly be known. They illustrate that this prediction can be outside the the convex hull of the individual calibrated predictions. The omniscient prediction, however, is not an aggregator as it is not a function of the predictions (or measurable with respect to $\F''$). Furthermore, it is always based on at least as much (and often more) information than the individual predictions combined. 
Therefore it is not clear how directly these results relate to forecast aggregation. Their work, however, did inspire us to look for a general inefficiency result.

In summary, existing general inefficiency results are limited to binary outcomes and the arithmetic mean of their calibrated predictions. 
Importantly, in each case the weights in the arithmetic mean are fixed ex-ante before the predictions are observed. Therefore the results do not apply to many common means such as the median or trimmed means. The discussion has also remained rather technical with no attempt to intuitively explain the reason or phenomenon behind the inefficiency. 
 

Part of our goal now is to settle this general conversation around inefficiency. First, we show that the inefficiency is not limited to the arithmetic mean of calibrated  predictions of a binary outcome but instead applies to almost all  means of calibrated predictions of any type of (univariate) outcome.
  Second, we intuitively explain the reasons behind this inefficiency. Third, to the best of our knowledge, we are the first to discuss the inefficiency of the means in the context of uncalibrated predictions.

\section{AGGREGATING CALIBRATED PREDICTIONS} \label{MeasuresOfCentralTendency}


\subsection{Arithmetic Mean} \label{contractionSection}

%
%
%
%


The (weighted) arithmetic mean is the most common way to combine predictions in practice \citep{hora2004probability, Ranjan08, jose2013trimmed}. 
  Therefore it is a natural starting point in our analysis of the means. 
  The arithmetic mean is defined and notated as follows.
\begin{class}[Arithmetic Mean] \label{Xw}
An arithmetic mean is $\mathcal{X}_w := \sum_{j=1}^N w_jX_j$, where each weight $w_j \geq 0$ is fixed and $\sum_{j=1}^N w_j = 1$. It is non-trivial if there exists a  pair $i \neq j$ such that $\P(X_i \neq X_j) > 0$ and $w_i, w_j > 0$.
\end{class}
Similarly to the previous studies mentioned in Section \ref{Lit Review}, the weights here cannot depend on the realized values of the predictions. Instead, they must be fixed before the predictions are observed. 
The non-triviality condition ensures that $\mathcal{X}_w$ does not assign all weight to a single prediction (or a group of a.s. identical predictions) but instead attempts to combine information from multiple different sources. Given that our goal is to analyze and understand aggregation of information from different sources, the discussion can safely focus only on the non-trivial cases.
That being said, the next theorem describes a non-trivial $\mathcal{X}_w$ in the light of the efficiency properties of Theorem \ref{optimal}. 


\begin{theorem}\label{contraction}
 If $\mathcal{X}_w$ is non-trivial, then the following holds. 
\begin{enumerate}[(i)]
\item  $\mathcal{X}_w$ is marginally consistent: $\E(\mathcal{X}_w) = \E(Y) = \mu_0$. \label{marginal calib Xw}

\item $\mathcal{X}_w$ is not calibrated: $\P\left[\E(Y | \mathcal{X}_w) \neq \mathcal{X}_w\right] > 0$.
\label{XwCalib}
  
 \item The variance $\Var(\mathcal{X}_w)$ is too low. More specifically, if $\mathcal{X}_w'' :=  \E(Y| \mathcal{X}_w)$ is the calibrated version of $\mathcal{X}_w$, then $\E(\mathcal{X}_w) = \E(\mathcal{X}_w'') = \mu_0$ but $\Var(\mathcal{X}_w) < \Var(\mathcal{X}_w'')$. In other words, $\mathcal{X}_w$ is under-confident in a sense that it is, in expectation, closer to the marginal mean $\mu_0$ than its calibrated version $\mathcal{X}_w''$ is. \label{underconfA}
 
\item $\mathcal{X}_w$ is not extremizing: $\Var(\mathcal{X}_w) < \max ( \Var(X_1), \dots, \Var(X_N))$. Thus $\Var(\mathcal{X}_w) < \Var(\mathcal{X}'')$. In other words,  $\mathcal{X}_w$ is under-confident in a sense that it is, in expectation, closer to the marginal mean $\mu_0$ than the efficient aggregator $\mathcal{X}''$ is.


\label{underconfB}

\item $\mathcal{X}_w$ is inefficient:
 $\P(\mathcal{X}_w \neq \mathcal{X}'') > 0$.
\end{enumerate}
\end{theorem}



Item (\ref{marginal calib Xw}) holds as long as each of the individual predictions is marginally consistent. The other items require calibrated predictions. To discuss these items, recall from Section \ref{outcomes} that calibration is equivalent to Assumption \ref{assumption2}. Item (\ref{XwCalib}) then shows that a non-trivial $\mathcal{X}_w$ is never an accurate assessment of any information set about $Y$ even though the individual predictions are. 
Instead it distorts the forecasters' information. 

This makes $\mathcal{X}_w$ under-confident in two different ways. 
First, item (\ref{underconfA}) defines under-confidence relative to its calibrated version $\mathcal{X}_w''$. Intuitively, $\mathcal{X}_w''$ represents an accurate assessment of all information left in $\mathcal{X}_w$. Given that $\mathcal{X}_w$ is closer to the non-informative prediction $\mu_0 = \E(Y|\F_0)$ than $\mathcal{X}_w''$ is, $\mathcal{X}_w$ is not as confident as it could be given all the information it has.
Second, item (\ref{underconfB}) defines under-confidence relative to $\mathcal{X}''$ and shows that $\mathcal{X}_w$ is not as confident as it should be, given the information it received through the predictions. 
Given that $\mathcal{X}_w$ is not extemizing or calibrated, it cannot use the forecasters' information efficiently under any probability space. This (strong) inefficiency is sometimes believed to arise because $\mathcal{X}_w$ represents the forecasters' shared information too many times \citep{kim2001inefficiency, palley2015eliciting}.  This, however, is not an entirely accurate explanation. In fact, the following example shows that a non-trivial $\mathcal{X}_w$ could benefit from representing shared information even more.
 For the reader's convenience, all technical derivations of the example have been deferred to the Appendix.

\begin{figure}[t]
\centering
\begin{minipage}[t]{.48\textwidth}
   \centering
\begin{tikzpicture}[->,>=stealth',shorten >=1pt,auto,node distance=3cm,
                    thick,main node/.style={circle,font=\sffamily\bfseries}]
        \clip (0,-0.3) rectangle (6, 4);
  \node[main node] (1) at (1,0) {$X_{(1)}$};
  \node[main node] (2) at (3.25,0) {$X_{(1,2)}$};
  \node[main node] (3) at (5.5,0) {$X_{(2)}$};
  \node[main node] (4) at (2.125,1.4) {$X_{1}$};
  \node[main node] (5) at (4.375,1.4) {$X_{2}$};
  \node[main node] (6) at (3.25,2.8) {$\mathcal{X}_w$};

  \path[every node/.style={font=\sffamily\small}]
    (1) edge node [left] {} (4)
    (2) edge node [right] {} (4)
     edge node [right] {} (5)
    (3) edge node [right] {} (5)
    (4) edge node [right] {} (6)
    (5) edge node [right] {} (6);
\end{tikzpicture}
   \caption{The information flow in Example \ref{intuition}. The variables $X_{(1)}, X_{(2)}$, and  $X_{(1,2)}$ represent the forecasters' private and shared information, respectively.}
   \label{HierarchyPlot}
\end{minipage}
\hspace{0.7em}
\begin{minipage}[t]{.48\textwidth}
   \centering
   \includegraphics[width = \linewidth]{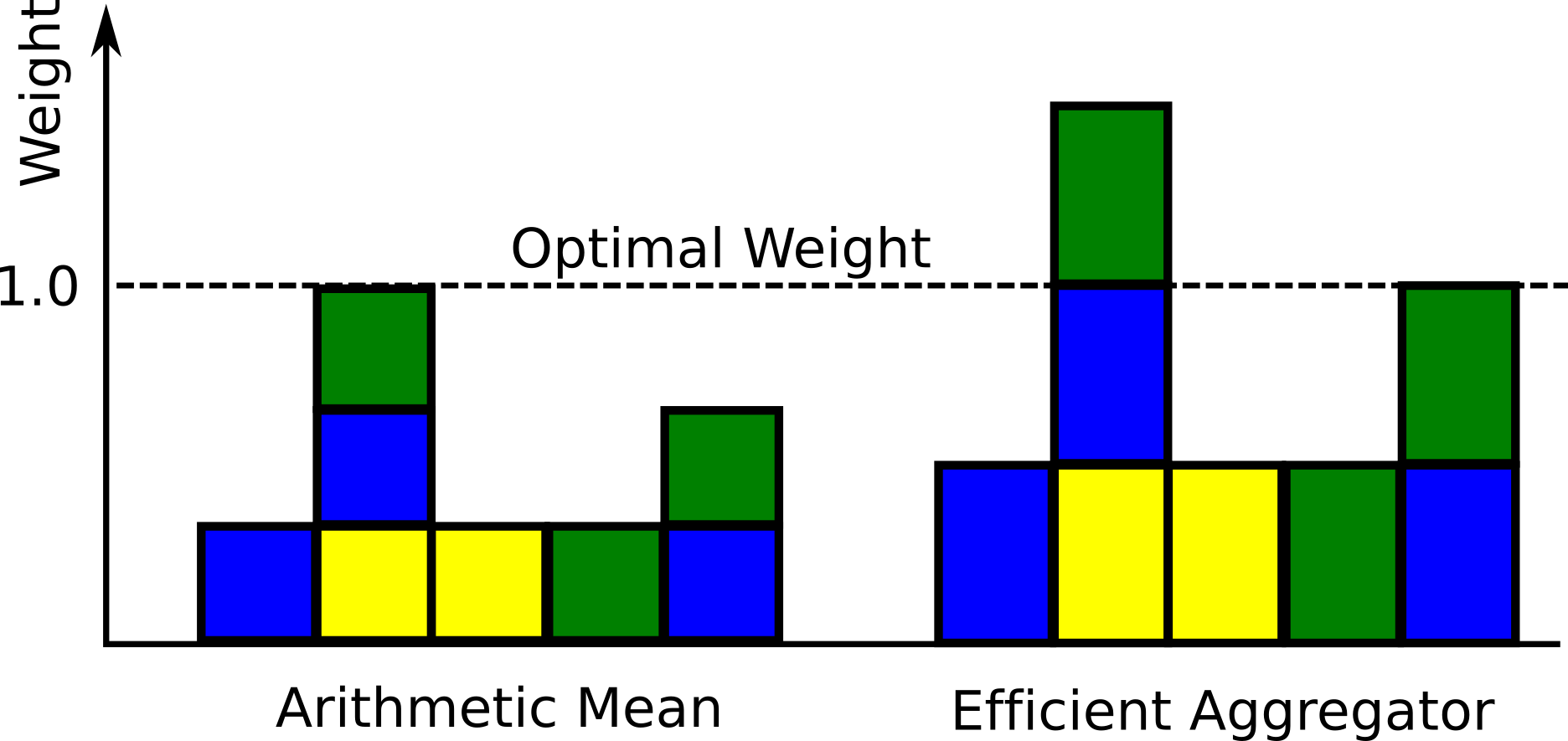} 
   \caption{Comparing the equally-weighted arithmetic mean and the efficient aggregator. Information consists of five pieces. From left to right, Yellow knows the second and third; Green knows the second and final two; Blue knows the last and first two.  }
   \label{Intuition}
\end{minipage}
\end{figure}

\begin{example}\label{intuition}
Consider $N = 2$ predictions of a continuous outcome $Y$. Suppose without loss of generality that $\E(Y) = \mu_0 = 0$.  
 The setup is described in Figure \ref{HierarchyPlot}. On the bottom level are three calibrated \textit{interns} predicting $Y$. Suppose their predictions $X_{(1)}, X_{(1,2)}$, and $X_{(2)}$ are independent. 
 On the middle level are two experts. Their private and shared information are determined by the interns' predictions. More specifically, for $j = 1,2$, the $j$th expert predicts $X_j = \E(Y|X_{(j)}, X_{(1,2)})$. This way $X_{(j)}$ represents the $j$th expert's private information and $X_{(1,2)}$ represents the experts' shared information. In the end, the experts report their predictions $X_1$ and $X_2$ to a decision-maker who combines them with the weighted average $\mathcal{X}_w = w_1X_1 + w_2X_2$. The objective is to explore how $\mathcal{X}_w$ represents the experts' private and shared information, namely $X_{(1)}$, $X_{(2)}$, and $X_{(1,2)}$. To make the aggregation task non-trivial, suppose $\Var\left(X_{(1)}\right), \Var\left(X_{(2)}\right) > 0$ such that $\P(X_1 \neq X_2) > 0$. 



%
%
%

The arithmetic mean $\mathcal{X}_w$ can be efficient only under some probability space where the efficient aggregator is linear, that is,  $\mathcal{X}'' = \beta_1X_1 + \beta_2X_2$ for some constants $\beta_1, \beta_2 \in \R$.  
If $\delta_j := \Var(X_j)$ and $\rho := \Cov(X_1, X_2)$, then minimizing the expected Bregman loss (such as the squared error loss) gives
 $\beta_j = (\delta_1\delta_2 - \rho \delta_i)/(\delta_1\delta_2 - \rho^2) \geq 0$, where $j,i \in \{1,2\}$ and $j \neq i$. This shows that if the experts' predictions are independent, then $\rho = 0$ and $\mathcal{X}'' = X_1 + X_2$. In other words, simple summing is the efficient way to combine information in independent predictions.  
 

Given that the interns' predictions are independent, the $j$th expert's prediction can be written as $X_j = X_{(j)} + X_{(1,2)}$. Plugging this into the aggregators gives
 \begin{align}
\mathcal{X}_w &= w_1 X_{(1)} + w_2 X_{(2)} + X_{(1,2)}  \text{ and }\label{X_wDecom}\\
\mathcal{X}'' &= \beta_1 X_{(1)} + \beta_2 X_{(2)} + (\beta_1+\beta_2)X_{(1,2)}, \label{X_ppDecom}
\end{align}
where $\beta_1 + \beta_2 > 1$ (see the Appendix for this result) such that $\P(\mathcal{X}_w \neq \mathcal{X}'') > 0$. 
This shows that $\mathcal{X}_w$ could be improved by relaxing its sum-constraint.

To understand why, observe that the optimal way to combine the three pieces of information, that is, the interns' predictions is via simple summing:
 \begin{align}
\mathcal{X}' := \E(Y|X_{(1)}, X_{(2)}, X_{(1,2)}) &= X_{(1)} + X_{(2)} + X_{(1,2)}.\label{X_pDecom}
\end{align}
This aggregator uses all information that exists in the example and therefore cannot be improved upon. Unfortunately, the decision-maker only has access to the experts' predictions and hence cannot separate out the three pieces of information in this manner.
  Therefore $\mathcal{X}'$ is not fully attainable in practice.  It does, however, offer a theoretical optimum, and every aggregator should aim to be as close to it as possible. 
  


 Comparing (\ref{X_pDecom}) to (\ref{X_wDecom}) shows that a non-trivial $\mathcal{X}_w$ weights the shared information exactly as it should but under-weights any private information.  
This shrinks the private information towards the non-informative prediction $\E(Y|\F_0) = \mu_0 = 0$ and makes $\mathcal{X}_w$ under-confident in comparison to $\mathcal{X}'$. If $\mathcal{X}_w$ was not sum-constrained, it could counter this under-confidence  by assigning more weight to each piece of information. This is precisely what $\mathcal{X}''$ does.

  Comparing  (\ref{X_pDecom}) to (\ref{X_ppDecom})  shows that $\mathcal{X}''$ cannot equal $\mathcal{X}'$ under any fixed $\beta_1$ and $\beta_2$. Therefore a linear aggregator cannot avoid over- or under-weighting some information. 
This forms a  trade-off between over-weighting shared and under-weighting private information.  The efficient aggregator balances this
%
in a way that brings it as close to $\mathcal{X}'$ as possible. In fact, $\mathcal{X}''$ is the orthogonal projection  of $\mathcal{X}'$ onto the space of linear aggregators. Even though $\mathcal{X}''$ loses due to over-weighting of the shared information, it is more than compensated for that loss by the increased  weights on the private information. 
Therefore, perhaps somewhat counter-intuitively, intentionally over-weighting shared information can be beneficial. 
\end{example}

This example illustrates a general challenge in forecast aggregation: private and shared information  in the predictions cannot be separated. Any weight given to $X_j$ is automatically assigned to both its private and shared components, leading to the shared-to-private information  tradeoff. 
Figure \ref{Intuition} illustrates this for $N = 3$ forecasters indexed by different colors. Each column represents a different piece of information. The height of a colored rectangle is the weight given to the corresponding forecaster, and the height of the entire column is the weight that is ultimately assigned to that piece of information. Observe how $\mathcal{X}_w$ assigns optimal weight only to information that is known by everyone (second column from the left); everything else is under-weighted. 
This is unfortunate because in many contexts the amount of information known precisely to everyone is likely to shrink as the group of forecasters grows. Consequently, $\mathcal{X}_w$ can end up underweighting a larger and larger share of the forecasters' information. 


On the flip side, this all suggests that $\mathcal{X}_w$ can work well when most of the forecasters' information is shared. A classic example comes from the 1906 country fair in Plymouth where 787 people guessed the weight of an ox \citep{galton1907vox}. By looking at the ox everyone assessed roughly the same information. Thus $\mathcal{X}_w$ should perform well. It in fact did! The equally weighted arithmetic mean was 1197 lbs., which is only one pound away from the true weight of 1198 lbs. There are similar results about estimating the number of jelly beans in a jar (average = 871, truth = 850), the temperature in a class room (average = 72.4 $^{\circ}$F, truth = 72.0 $^{\circ}$F), and others \citep{surowiecki2005wisdom}. In all these cases the forecasters are likely to use very similar information.

If the forecasters, however, use different information, $\mathcal{X}_w$ is likely to be under-confident. This was observed by the Good Judgment Project \citep{ungar2012good} that collected thousands of probability predictions about different future events, ranging from  international negotiations and economic shifts to military conflicts and global leadership. For instance, two of the events were ``Will Italy's Silvio Berlusconi resign, lose re-election/confidence vote, or otherwise vacate office before 1 January 2012?" and ``Will Moody's issue a new downgrade on the long-term ratings for any of the eight major French banks between 30 July 2012 and 31 December 2012?" Such prediction tasks are challenging and require much specific information. This is likely to cause high variation in the forecasters' information sets. Consequently, $\mathcal{X}_w$ should be under-confident. In fact, it was! Transforming the equally weighted arithmetic mean directly away from $0.5$ led to significant improvements in its squared error loss \citep{baron2014two, satopaa}.

\subsection{Means}\label{RandomSection}
Aside from the arithmetic mean, other common aggregators are the median, midrange, trimmed means, and other measures of central tendency. 
These can be expressed as arithmetic means but the weights would need to depend on the realized values of the predictions. Given that results in Section \ref{contractionSection} do not permit such random weights, a different more general analysis is required. We begin such an analysis with a formal definition of the mean.



\begin{class}[Means] \label{Xc}
A mean $\mathcal{X}_{[\cdot]}$ remains between the smallest and largest predictions, that is, $\mathcal{X}_{[\cdot]} \in [\min(X_1, \dots, X_N), \max(X_1, \dots, X_N)]$ a.s. The subindex reminds us that $\mathcal{X}_{[\cdot]}$ always stays within the closed convex hull. 
\end{class} 
This definition poses very little restrictions on the functional form of the aggregator. In fact, as long as the mean remains between the extreme predictions, it can decide how to treat the predictions before or after they have been reported. Unfortunately, the next example shows that this class is not entirely inefficient. 
\begin{example}\label{Example1}
Consider rolling a fair die and predicting the chances of an even number turning up. This can be described with the following probability space. Suppose $\Omega = \{1, \dots, 6\}$ and let $\F = 2^\Omega$ be all subsets of $\Omega$. The outcome is $Y = \mathbb{1}_A$, where $A = \{2, 4, 6\}$, and $\P$ is uniform over $\Omega$, that is, $\P(E) = |E|/6$ for all $E \in \F$.  Consider two forecasters with the information sets $\F_1 = \sigma(\{1\})$ and $\F_2 = \sigma(\{6\})$. 
For $\omega \in \Omega$ the relevant predictions are
\begin{align*}
X_1(\omega) &= \begin{cases}
0 & \text{ if } \omega = 1\\
3/5& \text{ otherwise} 
\end{cases},&
X_2(\omega) &= \begin{cases}
1 & \text{ if } \omega = 6\\
2/5 & \text{ otherwise} 
\end{cases},  &
\mathcal{X}'' &= \begin{cases}
0  & \text{ if } X_1 = 0 \\
1  & \text{ if } X_2 = 1\\
1/2 & \text{ otherwise} 
\end{cases}.
\end{align*}
\end{example}
This example shows that 
$\mathcal{X}_{[\cdot]}$ can be efficient if it is allowed to equal  the smallest or largest prediction when the forecasters disagree. If, however, Definition \ref{Xc} is tightened slightly such that this is not allowed, the class still contains essentially all common aggregators but inefficiency becomes a class property.

\subsection{Strict Means}\label{StrictConvex}
Motivated by Example \ref{Example1}, this section analyses the following class of aggregators. 
\begin{class}[Strict Means] \label{Xsc}
A strict mean $\mathcal{X}_{(\cdot)}$ equals the unanimous prediction if all predictions agree but  if at least two forecasters disagree, then $\mathcal{X}_{(\cdot)}$ remains strictly between the smallest and largest predictions, that is, $\mathcal{X}_{(\cdot)} \in (\min(X_1, \dots, X_N), \max(X_1, \dots, X_N)).$ The subindex reminds us that $\mathcal{X}_{(\cdot)}$ stays within the open convex hull when some forecasters disagree. 
\end{class} 
The strict mean is equivalent to the arithmetic mean with weights that can depend on the realized values of the predictions but under the constraint that positive weight is assigned to at least two different predictions when some forecasters disagree. 
This is still a very general class of aggregators. For one, given that measures of central tendency aim to summarize the predictions with a single central value, essentially all of them are strict means. Surely, it is possible to construct simple examples where this does not hold.  For instance, consider three predictions that can only take on two different values. Every time some of the predictions disagree, the median and mode equal the minimum or maximum prediction. This is admittedly a rather contrived example, and for most complex and interesting cases measures of central tendency can be assumed to be strict means. 

The generality of Definition \ref{StrictConvex} is further illustrated by the following enumeration of two common classes of aggregators that classify as strict means:
\begin{enumerate}[(i)]
\item Sometimes practitioners are advised to ``throw out the high and low predictions" \citep{armstrong2001combining} and use trimmed or Winsorized means instead \citep{jose2008simple}. All such aggregators are strict means.
\item The quasi-arithmetic mean (a.k.a., $f$-mean) is
$\mathcal{X}_{\Phi} := \Phi^{-1} \left[ \sum_{j=1}^N w_j \Phi(X_j) \right]$, where $\Phi$ is a strictly monotonic function from some interval $I \subseteq \R$ to $\R$, each  $w_j \geq 0$, and $\sum_{j=1}^N w_j = 1$ \citep{grabisch2011aggregation}. 
Popular members are the generalized means given by $\Phi(x) = x^{a}$ for some $a \in \R$. For instance, the harmonic, geometric, arithmetic, quadratic, and cubic means correspond respectively to $a = -1, 0, 1, 2, 3$. Quasi-arithmetic means with $\Phi(x) = \logit(x)$ or $\Phi(x) = \probit(x)$ are also popular in aggregation of probability predictions $X_j \in (0,1)$ \citep{satopaamodeling2}.  If all weights (random or not) are positive, these aggregators are strict means. 
\end{enumerate}

%


To analyze the efficiency of strict means, recall a well-known result stating that all $\sigma$-fields and hence all information sets $\F_j$ are either finite or uncountable \citep[Theorem 1.20]{davidson1994stochastic}. 
It turns out to be constructive to analyze strict means under each case separately. 
 %
%
%
%
%
%
%
%
First, we consider finite information sets. 


%
%


%
\begin{theorem}[Inefficiency of the Means]\label{centralTendency}
Suppose $|\F_j| < \infty$ for all $j = 1, \dots, N$.
 If at least two predictions disagree with positive probability, then
$\mathcal{X}_{(\cdot)}$ is inefficient. In other words, $\P(\mathcal{X}_{(\cdot)} \neq \mathcal{X}'') > 0$ as long as $\P(X_i \neq X_j) > 0$ for some pair $i \neq j$.\\
\vspace{-0.8em}\\
(\textit{Remark.} This result can be made somewhat more general because its proof only requires $\mathcal{X}_{(\cdot)}$ to remain within either pre-specified half-open convex hull of the predictions. Given that this minor generalization hardly adds anything to the discussion, it is only mentioned but excluded from the final theorem for the sake of clarity.)
\end{theorem}



Similarly to before, some forecasters are required to differ such that the problem is non-trivial and aggregation of information from different sources is necessary for achieving efficiency. 
Given how general Definition \ref{Xsc} is, detailed results akin to Theorem \ref{contraction} can be hardly expected. Therefore it is important to emphasize that Theorem \ref{centralTendency} simply states that $\mathcal{X}_{(\cdot)}$ does not use the forecasters' information efficiently as long as at least two forecasters differ. Thus $\mathcal{X}_{(\cdot)}$ distorts or leaves some information unused. Combining this with the efficiency properties in Theorem \ref{optimal} explains that if $\mathcal{X}_{(\cdot)}$  is calibrated, it is not extremizing (and \textit{vice versa}). 




%
%


Unfortunately, as the next example illustrates, inefficiency does not need to hold under uncountably large information sets. 

\begin{figure}[t]
   \centering
   \includegraphics[width = 0.6\linewidth]{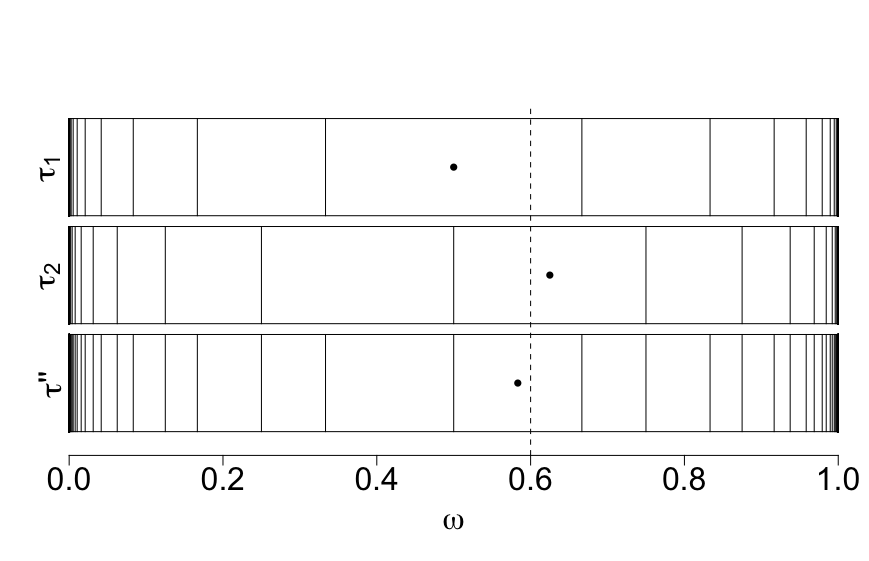} 
   \caption{The partitions $\tau_1$ and $\tau_2$ generate the forecasters' information $\F_1$ and $\F_2$. Together they form the partition $\tau''$ that generates the forecasters' combined information $\F''$. Under Lebesgue measure the predictions are the middle points of each interval. For instance, if $\omega = 0.6$ (the dashed line), then predictions are the points on the graph. Observe how $\mathcal{X}''$ is strictly between $X_1$ and $X_2$ under each $\omega \in [0,1]$.}
   \label{ExamplePartition}
\end{figure}

\begin{example} \label{uncountable}
Consider $N = 2$ forecasters predicting an unknown proportion that is uniformly distributed over the unit interval. This can be described with the standard probability space $(\Omega, \F, \P)$, where $\Omega = [0,1]$, $\F$ holds all the Borel subsets $\mathcal{B}([0,1])$, and $\P$ is the Lebesgue measure on $[0,1]$. The target proportion is defined as $Y(\omega) = \omega$ for all $\omega \in \Omega$. 

Suppose the forecasters are better at predicting proportions near zero and one.  To form their information sets, pick two strictly increasing sequences $\tau_1 = \{a_k : k \in \mathbb{Z}\}$ and $\tau_2 = \{b_k : k \in \mathbb{Z}\}$
such that both $a_k$ and $b_k$ converge to $0$ as $k \to -\infty$ and to $1$ as $k \to \infty$.
Suppose further that the sequences alternate in magnitude such that $a_{k} < b_{k} < a_{k+1} < b_{k+1}$ for all $k \in \mathbb{Z}$. For instance, one can let
\begin{align}
\tau_1 &= 
 \left\{\frac{1}{2} \pm \left( \frac{1}{6} - \frac{\gamma_k}{3} \right): k \in \mathbb{N}\right\} & \text{ and } && \tau_2 &=
 \left\{\frac{1}{2}, \frac{1}{2} \pm \frac{\gamma_k}{4} : k \in \mathbb{N}\right\}, \label{specificChoice}
\end{align}
where $\gamma_k = \sum_{j=0}^{k} \left( \frac{1}{2}\right)^j \to 2$ as $k \to \infty$. These sequences are illustrated in Figure \ref{ExamplePartition}.

Any such sequences $\tau_1$ and $\tau_2$ define two countably infinite partitions of $\Omega$. The atoms of these partitions accumulate only at the extremes, and no atom in one partition is contained within an atom in the other partition.
The $j$th forecaster's information set $\F_j$ is then generated from the partition defined by $\tau_j$. This computes all possible unions of the atoms. The result is an uncountably infinite information set because it has the same cardinality as the power set of a countably infinite set. 
 Similarly, the forecasters' combined information $\F''$ is generated by the partition
$\tau'' = \tau_1 \cup \tau_2 = \{\dots, a_{-1}, b_{-1}, a_0, b_0, a_1, b_1, \dots\}$. 

The predictions $X_1$, $X_2$, and $\mathcal{X}''$ equal the middle points of those atoms to which the realized $\omega \in \Omega$ belongs. In other words, if $\omega \in [a_j, a_{j+1}] \cap [b_i, b_{i+1}]$, then 
\begin{align}
X_1 &= (a_j+a_{j+1})/2, &
X_2 &= (b_i+b_{i+1})/2, &
\mathcal{X}'' &= \begin{cases}
(b_i+a_{j+1})/2 & \text{ if } a_j < b_i\\
(b_{i+1}+a_{j})/2 & \text{ if } a_j > b_i
\end{cases}. \label{Example2Rev}
\end{align}
Given the way the values of $\tau_1$ and $\tau_2$ alternate, $\mathcal{X}''$ is always strictly between $X_1$ and $X_2$. This shows that $\mathcal{X}_{(\cdot)}$ (and hence $\mathcal{X}_{[\cdot]}$) can be efficient if the forecasters' information sets are allowed to be uncountably large. 

Under the specific sequences (\ref{specificChoice}), the efficient aggregator (\ref{Example2Rev}) becomes
\begin{align*}
\mathcal{X}''
&= \begin{cases}
\frac{2}{3} \min(X_1,X_2) + \frac{1}{3} \max(X_1,X_2) & \text{ if } X_2 < 0.5\\
\frac{1}{3} \min(X_1,X_2) + \frac{2}{3} \max(X_1,X_2) & \text{ if } X_2 > 0.5\\
\end{cases}.
\end{align*}
This is an arithmetic mean but the weights depend on the realized values of the predictions. In particular, a weight of $2/3$ is given to the prediction that is closer to the nearest extreme value (at $0.0$ or $1.0$). Compared to the equally weighted arithmetic mean $(X_1+X_2)/2$, the aggregator $\mathcal{X}''$ is further away from the marginal mean $\mu_0 = \E(X_j) = 0.5$ and hence more confident. 

\end{example}

Given that the inefficiency of the means (Theorem \ref{centralTendency}) does not need to hold under uncountably large information sets, it is natural to ask whether such sets even exist in the real world. 
The answer is ``no'' because information content is finite in the physical world \citep{hibbard2014self}. In fact, according to \cite{lloyd2002computational}, the observable universe has a finite information capacity of at most $10^{120}$ bits. Even if the universe did contain infinite information, forecasters could not process it. Computers have finite memory and represent the world in a discrete form. 
Similarly, the human cognition is limited in its ability to process information.
Therefore, even though a forecaster with infinite information may be a convenient approximation in some theoretical work, it is not a precise description of reality. \cite{casella2002statistical} even argue that in practice $|\Omega| < \infty$ (and hence $|\F_j| < \infty$ for all $j$) because measurements cannot be made with infinite precision. Either way, we conclude that (strict) means can be efficient aggregators of calibrated predictions in theory but not in practice.
%
%


It may be that under some additional restrictions on the predictions and outcome the entire class of means is inefficient. Given, however, that we begun with the most general hypothesis  ``all means are inefficient'' and then justified each necessary specification with a concrete example, we believe that Theorem \ref{centralTendency} is very close, if not equal, to the most general version of the inefficiency of the means.

\section{AGGREGATING NOISY PREDICTIONS} \label{noisy} 

\subsection{Asymptotic Efficiency}
So far our results have depended on the predictions being calibrated. 
This section, however, shows that the results can shed light even upon contexts where predictions are not calibrated. The lack of calibration here is achieved by adding noise to the calibrated predictions. This way each prediction can be described as a distribution with the center at the calibrated prediction. If these noisy predictions are aggregated with a similar notion of centrality, then intuitively the aggregator should converge to a central point of the calibrated predictions and hence not be (asymptotically) efficient.

To verify this intuition, denote the $j$th noisy prediction with
\begin{align}
\tilde{X}_j = Q\left[ \E(Y|\F_j), \epsilon_j \right], \label{pif model}
\end{align}
where $Q(\cdot, \cdot)$ is a deterministic function that adds i.i.d. error $\epsilon_j$ to the calibrated prediction $\E(Y|\F_j)$. For instance, if $Y$ is real-valued, one can simply let $Q(x,y) = x+y$, giving $\tilde{X}_j = \E(Y|\F_j) + \epsilon_j$.

To conduct asymptotic analysis, we must describe how the group of forecasters grows.
First, recall from Section \ref{StrictConvex} that in practice $|\Omega| < \infty$. The principal $\sigma$-field $\F$ is then finite and all of its sub-$\sigma$-fields can be collected into a finite set  $H = \{\H_1, \H_2, \dots, \H_I\} = \{\sigma(\H) : \H \subseteq \F\}$. 
Let the $\F_j$s be i.i.d. draws from $H$ with $p_i := \P(\F_j = \H_i)$. 
%
 Now, by Borel-Cantelli II \citep[Theorem 2.3.6.]{durrett2010probability}, $\sigma(\F_1, \F_2, \dots) \stackrel{a.s.}{\to} \sigma( \H_i : p_i > 0)$ as $N \to \infty$.
%
%
An aggregator $\mathcal{X}$ is then asymptotically efficient if and only if $\mathcal{X} \stackrel{a.s.}{\to} \E[Y | \sigma( \H_i : p_i > 0)]$.

Now, denote the cumulative distribution function (CDF) of $\tilde{X}_j | (\F_j = \H_i)$ with $\tilde{F}_i$. The noisy predictions are then i.i.d. draws from the mixture distribution $\tilde{F} = \sum_{i=1}^I p_i \tilde{F}_i$. Describe the notion of centrality with a statistical functional $T(\cdot)$ that $T(\tilde{F}_i) = \E(Y|\H_i)$.
 In words, $T$ recovers the calibrated prediction from the distribution of its noisy version. 
Lastly, denote the support of a random variable $X$ with $\supp(X) = \{ t: \P(X = t) > 0\}$. With this notation, assume the following:
\begin{assumption}
$p_i, p_{i'} > 0$ and $\P[ \E(Y|\H_i) \neq \E(Y|\H_{i'})] > 0$ for some $i \neq i'$. \label{Assump5}
\end{assumption}
\begin{assumption}
Let $X \sim F$ and $X' \sim F'$.  If $\P(X \leq t) > \P(X' \leq t)$ for all $t \in \supp(X)$, or $\P(X' \geq t) > \P(X \geq t)$ for all $t \in \supp(X')$, then $T(F) < T(F')$ \label{Assump3}
\end{assumption}
\begin{assumption}
If $x < y$,  then  $Q(x,\epsilon) < Q(y,\epsilon)$ for all values of $\epsilon$. \label{Assump4}
 \end{assumption}
These conditions are  intuitive and rather mild. Assumption \ref{Assump5} ensures that the aggregation problem is non-trivial. Assumption \ref{Assump3} states that moving density of a distribution uniformly to one direction (left or right) strictly changes the functional value in the same direction. 
Assumption \ref{Assump4}, on the other hand, says that the prediction increases if the forecaster ``aims'' higher but keeps the same level of error. 

This all leads to the following corollary. 
\begin{corollary} \label{asympEff}
Consider an aggregator that $\mathcal{X} \stackrel{a.s.}{\to} T(\tilde{F})$. If assumptions (\ref{Assump5}) to (\ref{Assump4}) hold, then $\mathcal{X}$ 
converges a.s. to a strict mean of the $\E(Y|\H_i)$s and hence is not asymptotically efficient. 
\end{corollary}



Corollary \ref{asympEff} applies only if the aggregator $\mathcal{X}$ is strongly consistent for $T(\tilde{F})$. This holds, for instance, if $\mathcal{X} = T(\hat{F})$, where $\hat{F}$ is the empirical CDF of the $\tilde{X}_j$s, and $T$ is Lipschitz in a sense that $|T(F) - T(G)| \leq K ||F-G||_\infty$, where $0 < K < \infty$ and $||F-G||_\infty = \sup_x|F(x)-G(x)|$ is the Kolmogorov metric \citep[Exercise 2.7.5, p. 24]{wassermann2006all}. 
 As long as the forecasters do not always either under- or overestimate the calibrated prediction, this plug-in estimator $T(\hat{F})$ can be seen to classify as a mean. 

To conclude, we illustrate Corollary \ref{asympEff} under the arithmetic mean $\mathcal{X}_w$ and a simple model of noisy predictions. 
 This assumes that $\tilde{X}_j = \E(Y|\F_j) + \epsilon_j$, where the $\epsilon_j$s  are i.i.d. with $\E(\epsilon_j) = 0$. Thus $T(F)$ is the expected value. 
Let $w_j = b_j / B_N$, where $b_1, b_2, \dots$ is a sequence of positive values and $B_N := \sum_{j=1}^N b_j \to \infty$ as $N \to \infty$. This way each $w_j$ is positive but decreases in $N$. Now, consider the counting function 
$\gamma(t) = \#\{N \geq 1 : B_N / b_N \leq t\}$.
  Then $\mathcal{X}_w \stackrel{a.s.}{\to} T(\tilde{F})$  if and only if  $\limsup_{t \to \infty} \gamma(t) / t < \infty$ \citep{jamison1965convergence}. For instance, if $w_j = 1/N$ for all $j =1, \dots, N$, then $\gamma(t)  = \lfloor t \rfloor$ and $\limsup_{t \to \infty} \gamma(t) / t  = 1$, returning the strong law of large numbers \citep[Theorem 2.4.1., p. 73]{durrett2010probability}. If this condition holds in our example, then $\mathcal{X}_w \stackrel{a.s.}{\to} T(\tilde{F}) = \sum_{i=1}^I p_i \E(Y|\H_i)$. This means that, by assumption (\ref{Assump5}), $\mathcal{X}_w$ converges a.s. to a non-trivial arithmetic mean of the $\E(Y|\H_j)$s and hence, by Theorem \ref{contraction}, is not asymptotically efficient (or even calibrated or extremizing).

This example suggests a crucial conflict between means and information aggregation. The arithmetic mean $\mathcal{X}_w$ converges to a typical (defined here in terms of the expected value) calibrated prediction. Statisticians often use means as references for the typical \citep{weisberg1992central}. Information aggregation, however, is not about estimating a typical prediction; it is about estimating the calibrated prediction that a forecaster with all of the group's information would make. This prediction, of course, does not have to be typical. In fact, the individual predictions may not ever come close to it. This way information aggregation and summarizing the distribution of a sample with a typical value can be seen to be very different exercises. 

\section{DISCUSSION} \label{conclusion}

Since its beginning statistics has been about studying variability \citep{weisberg1992central}. Of particular interest is to understand how individual observations vary around some unknown quantity of interest. The statistician then models this variability and derives an aggregator that can combine the observations into an accurate estimate of the target quantity. 

Historically the most common model of variability can be illustrated with a sensitive instrument that repeatedly estimates some target quantity. Here low variability is associated with better accuracy. Deviations represent idiosyncratic error, and the estimates are treated as draws from a distribution with the center at the target quantity. Depending how this center is defined, the target quantity is often estimated with some mean of the observations. The idea is that  through aggregation the errors in the ``too high'' estimates will cancel out the errors in the ``too low'' estimates, leaving a low error estimate of the target quantity. This way means can be seen as mechanisms for error reduction \citep{plackett1958studies}. 


As the world faces increasingly complex forms of data, existing models of variability must be occasionally revisited and any shortcomings addressed. For instance, estimates do not always come from a single instrument anymore. The current crowdsourcing technology allows decision-makers to easily and cheaply ask a large number of people for their predictions about some future event. These forecasters may 
 have different information, making their predictions very different. Of course, they may make errors in estimating their ideal predictions but such errors represent lack of skill -- a concept that is very different from information (see, e.g., \citealt{chamorro2015wiley}).
Here it is not enough to just reduce error. Instead the aggregator must be able to make sense of the information among the forecasters and combine it into a consensus prediction. 

This paper showed that (strict) means cannot make sense of the forecasters' information even if the forecasters accurately assess their own information and hence report calibrated predictions. 
To explain this intuitively, recall from Section \ref{InfoCollection} that efficient aggregators of calibrated predictions associate higher variability with higher levels of information content. These aggregators aim to capture all of the forecasters' information into a consensus that is more informed and hence more variable than the individual predictions. They hence have a tendency towards the extremes. Means, however, do the opposite: they tend towards a central location and aim to reduce variability that is perceived as lack of accuracy. This contrasting tendency and opposing interpretation of variability is an easily understood explanation of the inefficiency of the means.

If essentially all means are inefficient, a natural question to ask is whether 
an efficient aggregator of calibrated predictions even exists in practice. 
Fortunately, the answer is ``yes.'' One solution is to assume a parametric model for calibrated predictions and then find 
 $\mathcal{X}''$ by either analytically  or  numerically under the chosen model.
Of course, choosing a probability model in a given context requires a certain amount of field expertise. If such expertise is not available, one can work with 
a generic model such as the \textit{Gaussian partial information model} \citep{satopaamodeling2}. This model brings the partial information framework (Section \ref{outcomes}) to practice by only assuming Gaussianity -- nothing else. 


In practice, however, forecast variation is likely to be driven by error and information, like in (\ref{pif model}). On one hand, means can reduce error but not aggregate information (Section \ref{noisy}). On the other hand,  aggregators based on the partial information framework interpret all variation as information. Consequently, they will overfit to error. This motivates some of our current work-in-progress to develop an aggregator that mimics the behavior of a smoothing spline: the aggregate is a compromise between error and information. 
That is, depending on what portion of the variability is due to error, the aggregator can take on different forms between a mean and a partial information aggregator. 
 The preliminary results look promising but much more development and testing is needed. 

The purpose of this paper was not to criticize the existing literature on the aggregation properties of the means (see, e.g., \citealt{lichtendahl2013wisdom}).
Instead our goal is to direct researchers' attention towards the type of variability that is present among predictions and hence inspire new research in forecast aggregation.

 In terms of practice, we hope that care is taken in matching the aggregator with the forecasting context. Our work explains that error and information motivate very different forms of aggregation. Applying one type of aggregation to a wrong type of forecast variability yields sub-optimal performance. As an analogy, this is like making conclusions based on a linear regression model when the true underlying relationship is not linear. One must always consider how the assumptions behind the statistical techniques align with the data. This is no different when it comes to aggregating predictions.

\appendix
\section{APPENDIX: PROOFS AND TECHNICAL DETAILS}
\subsection{Proof of Theorem \ref{optimal}}
$\Rightarrow$ Assume that $\mathcal{X}''$ is efficient.
\begin{enumerate}
\item[(i)] Given that $ \mathcal{X}'' = \E(Y | \F'')$, we have $\sigma(\mathcal{X}'') \subseteq \F''$. The law of total expectation then gives 
$\E(Y | \mathcal{X}'')
= \E[\E(Y|\F'')|\mathcal{X}'']
= \E(\mathcal{X}''|\mathcal{X}'')
= \mathcal{X}''$.


\item[(ii)] Let $\mathcal{X}_\v = \E(Y | X_j, j \in \v)$ for some subset $\v \subseteq \{1, \dots, N\}$.  Then $\sigma(\mathcal{X}_\v) = \F_\v \subseteq \F'' = \sigma(X_1, \dots, X_N)$. 
By \citet[Proposition 2.1.]{satopaamodeling2}, item (i), and the Cauchy-Schwarz inequality, we have that $\Var(\mathcal{X}_\v) = \Cov(\mathcal{X}_\v, \mathcal{X}'')\leq \sqrt{\Var(\mathcal{X}'') \Var(\mathcal{X}_\v)  }$.
Squaring and diving both sides by $\Var(\mathcal{X}_\v) $ gives the final result. 
\end{enumerate}

\noindent
$\Leftarrow$ Assume that an aggregator $\mathcal{X}$ satisfies properties (i) and (ii). By definition of the aggregator, $\mathcal{X} \in \F''$ such that $\G := \sigma(\mathcal{X}) \subseteq \F''$. Then $\mathcal{X} = \E(Y|\G) = \E[\E(Y|\F'')|\G] = \E(\mathcal{X}''|\mathcal{X})$. 
Thus $\mathcal{X}$ is also calibrated for $\mathcal{X}''$. Now, by property (ii), $\Var(\mathcal{X}'') \leq \Var(\mathcal{X})$. But given that $\G \subseteq \F''$, $\Var(\mathcal{X}) \leq \Var(\mathcal{X}'')$ \citep[Proposition 2.1.]{satopaamodeling2}. Thus $\Var(\mathcal{X}'') = \Var(\mathcal{X}) \Leftrightarrow \E\left[\mathcal{X}''^2\right] = \E\left[\mathcal{X}^2\right]$. From \citealt[Exercise 5.1.11]{durrett2010probability}, it follows that $\mathcal{X} = \mathcal{X}''$ a.s.

\qed

\subsection{Proof of Theorem \ref{contraction}}
\begin{enumerate}[(i)]
\item
Let $\w = (w_1, \dots, w_N)'$ and $\X = (X_1, \dots, X_N)'$. Then,
$\E(\mathcal{X}_w) = \E(\w' \X) = \w' \E(\X) = \mu_0 \w' \one_N = \mu_0$.

\item This and the next item are are generalizations of the proof in \cite{Ranjan08} who consider calibrated predictions of a binary outcome. 

Proof by contradiction: Suppose that a non-trivial $\mathcal{X}_w$ is calibrated. For any calibrated prediction (aggregate or individual) $X$,  we have that $ \E[(Y-X)^2] =  \E(Y^2)-\E(X^2)$. 
For $\mathcal{X}_w$ this is
\begin{align*}
 &\hspace{1.3em} \E\left[\left(Y-\w'\X\right)^2\right] = \E\left\{\left[\sum_{j=1}^Nw_j (Y-X_j)\right]^2\right\} \\
&= \sum_{i=1}^N\sum_{j=1}^Nw_iw_j \E\left(Y^2-YX_i-YX_j+X_jX_i\right) \\
&= \sum_{i=1}^N\sum_{j=1}^Nw_iw_j \E\left[ \E\left( Y^2|X_i \right)-\E\left( YX_i|X_i \right)-\E\left( YX_j|X_j\right)+X_jX_i \right] \\
&= \sum_{i=1}^N\sum_{j=1}^Nw_iw_j \E\left[\E\left(Y^2|X_i\right) -X_jX_i-\left(X_i-X_j\right)^2\right] \\
&= \sum_{i=1}^N\sum_{j=1}^Nw_iw_j \E\left[\E\left(Y^2|X_i\right) -X_jX_i\right]-\sum_{i=1}^N\sum_{j=1}^Nw_iw_j\E\left[\left(X_i-X_j\right)^2\right] \\
&= \left[\E\left(Y^2\right) -  \E\left(\mathcal{X}_w^2\right) \right]-\sum_{i=1}^N\sum_{j=1}^Nw_iw_j\E\left[\left(X_i-X_j\right)^2\right].
\end{align*}
The double sum on the final line is strictly positive because $\mathcal{X}_w$ is non-trivial. This means that $\mathcal{X}_w$ cannot be calibrated, giving us a contradiction.
Thus $\P[\mathcal{X}_w \neq \E(Y|\mathcal{X}_w)] > 0$. 

\item Given that $\mathcal{X}_w''$ is calibrated, $\E(\mathcal{X}_w'') = \mu_0$ and $\E[(Y-\mathcal{X}_w'')^2]  = \E(Y^2)-\E(\mathcal{X}_w''^2)$. Item (ii) shows that $\mathcal{X}_w$ is not calibrated. Thus $\P(\mathcal{X}_w'' \neq \mathcal{X}_w) > 0$ such that $\E[(\mathcal{X}_w'' - \mathcal{X}_w)^2] > 0$. This all gives us
\begin{align*}
&\hspace{1.3em}   \E \left[ \left( Y - \mathcal{X}_w\right)^2\right]
= \E \left[ Y^2 + 2\left(\mathcal{X}_w''^2-\mathcal{X}_w''^2\right)- 2Y\mathcal{X}_w + \mathcal{X}_w^2\right]\\
&= \E\left[\left(Y-\mathcal{X}_w''\right)^2\right] + \E\left[\left(\mathcal{X}_w-\mathcal{X}_w''\right)^2\right] 
= \E(Y^2)-\E(\mathcal{X}_w''^2) + \E\left[\left(\mathcal{X}_w-\mathcal{X}_w''\right)^2\right]\\
&> \E(Y^2)-\E(\mathcal{X}_w''^2).
\end{align*}
Furthermore, from the proof of item (ii), $\E \left[ \left( Y - \mathcal{X}_w\right)^2\right] < \E(Y^2)-\E(\mathcal{X}_w^2)$. Combining this with the above inequality gives the final result:
\begin{align*}
&&\E(Y^2)-\E(\mathcal{X}_w''^2)  &< \E(Y^2)-\E(\mathcal{X}_w^2)\\
\Leftrightarrow && \E(\mathcal{X}_w''^2) - \mu_0^2  &> \E(\mathcal{X}_w^2) - \mu_0^2\\
\Leftrightarrow && \Var(\mathcal{X}_w'')  &> \Var(\mathcal{X}_w^2).
\end{align*}



\item First, consider two calibrated predictions $X_i$ and $X_j$. If $\Corr(X_i, X_j) = 1$, then $X_i$ and $X_j$ are linearly dependent and $\F_i = \sigma(X_i) = \sigma(X_j) = \F_j$. Thus, it must be the case that $X_i = X_j$ a.s.  On the other hand, if $X_i = X_j$ a.s., then clearly $\Corr(X_i, X_j) = 1$. Thus $\Corr(X_i, X_j) = 1$ if and only if $X_i = X_j$ a.s. Now, let $m = \argmax_j  \Var(X_j) $ identify the prediction with the maximal variance.  Given that $\mathcal{X}_w$ is non-trivial, there is some pair $i \neq j$ such that $\P(X_i \neq X_j) > 0$ and $w_i,w_j > 0$. For these forecasters, $\Corr(X_i \neq X_j) < 1$ such that $\Cov(X_i, X_j) < \sqrt{\Var(X_i)\Var(X_j)} \leq \Var(X_m)$. Now, given that the maximum element in $\Cov(\X)$ is $\Var(X_m)$ and $\sum_{i,j} w_i w_j = 1$, we have that  
$\Var(\mathcal{X}_w) 
= \sum_{i,j} w_iw_j \Cov(X_i,X_j)
< \Var(X_m)$.

\item Proof by contradiction: Assume that $\mathcal{X}_w$ is efficient such that $\mathcal{X}_w = \mathcal{X}''$ a.s.  Property (iv), however, gives a contradiction. Thus $\P(\mathcal{X}_w \neq \mathcal{X}'') > 0$. 
\end{enumerate}
\qed


\subsection{Technical Details of Example \ref{intuition}}
Let $\X = (X_1, \dots, X_N)'$. If $\E(X_j) = \E(Y) = 0$ and $\Cov(\X) := \bSigma$,  it is well-known that $\argmin_{\boldsymbol{b} \in \R^N} \E(Y- \boldsymbol{b}'\X)^2 =  \Cov(\X, Y) \bSigma^{-1} =: \bBeta$ , where 
$\Cov(\X, Y) = (\Cov(X_1,Y), \dots, \allowbreak \Cov(X_N,Y)).$ \cite{satopaamodeling2} show that if $X_j$ is calibrated, then $\Cov(X_j, Y) = \Var(X_j)$ and hence $\bBeta = \diag(\bSigma)\bSigma^{-1}$. If $\Cov(X_j,X_i) = 0$ for all $i \neq j$, then $\bSigma = \diag(\bSigma)$, $\bBeta = \one_N'$, and $\bBeta \X = \sum_{j=1}^N X_j$. Thus information from independent sources combines via simple summing. 

If $N=2$, $\delta_j = \Var(X_j)$, and $\rho = \Cov(X_1, X_2)$, then $\bBeta = (\beta_1, \beta_2) =  \diag(\bSigma)\bSigma^{-1} = (\delta_1\delta_2 -\rho\delta_2, \delta_2\delta_2 -\rho\delta_1) / (\delta_1\delta_2 - \rho^2)$.
Given that $\delta_ j = \Var(X_{(j)}) + \Var(X_{(1,2)}) > \Var(X_{(1,2)}) = \rho$, it follows that $\rho < \min(\delta_1,\delta_2)$ and hence that $\beta_j > 0$. Now, $\beta_1 + \beta_2 = (2\delta_1\delta_2 - \rho\delta_1 - \rho\delta_2) / (\delta_1\delta_2 - \rho^2)$
and
\begin{align*}
\frac{\partial}{\partial \rho} (\beta_1 + \beta_2) &= - \frac{\delta_2(\delta_1-\rho)^2 + \delta_1(\delta_2-\rho)^2}{(\rho^2 - \delta_1\delta_2)^2} < 0.
\end{align*}
Thus the sum $\beta_1 + \beta_2$ is a strictly decreasing function of $\rho$.  Suppose without loss of generality that $\delta_1 \leq \delta_2$. Then, 
\begin{align*}
\beta_1 + \beta_2 &> (\beta_1 + \beta_2) |_{\rho = \delta_1} = \frac{2\delta_1\delta_2 - \delta_1^2 - \delta_1\delta_2}{\delta_1\delta_2 - \delta_1^2} = 1.
\end{align*}

Observe that $\Var(\mathcal{X}_w - \mathcal{X}'' ) = (w_1-\beta_1)^2 \Var(X_{(1)}) + (w_2-\beta_2)^2 \Var(X_{(2)}) + (1-\beta_1-\beta_2)^2 \Var(X_{(1,2)})$. As $\beta_1 + \beta_2 > 1$, we have that $w_1 \neq \beta_1$ or $w_2 \neq \beta_2$. Given that $ \Var(X_{(1)}) > 0$ and $\Var(X_{(2)}) > 0$, we have that $\Var(\mathcal{X}_w - \mathcal{X}'' )  > 0$ and hence $\P(\mathcal{X}_w \neq \mathcal{X}'' ) > 0$. 

Finally, to show that $\mathcal{X}''$ is the orthogonal projection of $\mathcal{X}'$ onto the space of all aggregators of the form $\beta_1 X_1 + \beta_2 X_2$, recall that $\argmin_{\boldsymbol{b} \in \R^N} \E(Y- \boldsymbol{b}'\X)^2 =   \Cov(\X, \mathcal{X}') \bSigma^{-1}$. Now, 
$\Cov(X_j, \mathcal{X}') = \Cov(X_{(j)} + X_{(1,2)}, X_{(j)} + X_{(i)} + X_{(1,2)}) = \Cov(X_{(j)}, X_{(j)}) + \Cov(X_{(1,2)},X_{(1,2)}) \allowbreak= \Var(X_j).$
Thus $ \Cov(\X, \mathcal{X}') \bSigma^{-1} = \diag(\bSigma)\bSigma^{-1}$ as desired.

\subsection{Proof of Theorem \ref{centralTendency}}
Proof by contradiction: 
Suppose the efficient aggregator $\mathcal{X}''$ is a.s. a strict mean. Recall from Section \ref{outcomes} that $\F_j = \sigma(X_j) = \{\{\omega \in \Omega : X_j(\omega) \in B\} : B \in \mathcal{R}\}$. Given that the Borel $\sigma$-field $\mathcal{R}$ is countably generated \citep[Exercise 1.1.4., p. 9]{durrett2010probability}, $\F_j$  is also countably generated \citep[Exercise 1.3.1., p. 16]{durrett2010probability}. Furthermore, as $|\F_j| < \infty$, there is a finite partition $\mathcal{A}_j = \{A_{1j}, A_{2j}, \dots, A_{n_jj}\}$ of $\Omega$ such that $\F_j = \sigma(\mathcal{A}_j)$ for all $j = 1, \dots, N$.
 The forecasters' combined information is $\F'' = \sigma\left( \cup_{j=1}^N \F_j \right) =\sigma(\mathcal{B}'')$, where $\mathcal{B}'' := \{B_1, \dots, B_{n''}\} =  \{B : B \neq \emptyset, B = \cap_{j=1}^N A_{j}, A_{j} \in \mathcal{A}_j\}$ is a partition of $\Omega$. This partition is finite $n'' < \infty$ because $n_j < \infty$ for all $j = 1, \dots, N$. 
Given that each $B \in \mathcal{B}''$ is contained in only one atom per $\mathcal{A}_j$, the $X_j$s and $\mathcal{X}''$ remain constant within each atom of $\mathcal{B}''$. 
If $X_j$ or $\mathcal{X}''$ is constant over some atom $C$, then denote the prediction with  $X_j(C)$ or $\mathcal{X}''(C)$, respectively.
Lastly, define $\mathcal{B}_+'' := \{B \in \mathcal{B}'' : \P(B) > 0\}$.



By construction, each $A_{ij} \in \mathcal{A}_j$ can be written as a unique union of some atoms in $\mathcal{B}''$.  Furthermore, given that $X_j = \E(Y|\F_j)$, we have, for any atom $A_{ij} \in \mathcal{A}_j$ such that $\P(A_{ij}) > 0$,
\begin{align}
 \int_{A_{ij}} Y d\P &= \int_{A_{ij}} X_j d\P = X_j(A_{ij}) \int_{A_{ij}}  d\P  = X_j(A_{ij}) \P(A_{ij}) \nonumber\\ 
\Leftrightarrow X_j(A_{ij}) &= \frac{\int_{A_{ij}} Y d\P}{\P(A_{ij})} = \frac{\sum_{k : B_k \cap A_{ij} \neq \emptyset}\int_{B_k} Y d\P}{\P(A_{ij})} = \frac{\sum_{k : B_k \cap A_{ij} \neq \emptyset}\mathcal{X}''(B_k) \P(B_k) }{\P(A_{ij})}\nonumber\\
\Leftrightarrow X_j(A_{ij}) &= \sum_{k : B_k \cap A_{ij} \neq \emptyset}p_{ijk}\mathcal{X}''(B_k), \label{decomposition}
\end{align}
where $p_{ijk} = \P(B_k) / \P(A_{ij})$ and $\sum_{k : B_k \cap A_{ij} \neq \emptyset}p_{ijk} = 1$. This expresses every value of $X_j$ as a convex combination of some values of $\mathcal{X}''$. 

Given that the $X_j$s and $\mathcal{X}''$ are integrable, they must be finite over all atoms $B \in \mathcal{B}_+''$. Suppose for now that there is a pair $i \neq j$ such that $X_i(B) \neq X_j(B)$ for some $B \in \mathcal{B}_+''$.  We can then define two different indices ${k}^* := \argmin_j\{X_j(B)\}$ and $k^{**} := \argmax_j\{X_j(B)\}$.
Given that, over this atom, the predictions do not all agree, the strict mean $\mathcal{X}''$ is in the interior of their convex hull:
\begin{align*}
X_{k^*}(B) &< \mathcal{X}''(B) < X_{k^{**}}(B)\\
\Leftrightarrow \sum_{k : B_{k} \cap A_{ik^*} \neq \emptyset, B \cap A_{ik^*} \neq \emptyset}p_{ik^*k}\mathcal{X}''(B_{k}) &< \mathcal{X}''(B)  <    \sum_{k : B_{k} \cap A_{ik^{**}} \neq \emptyset, B \cap A_{ik^{**}} \neq \emptyset}p_{ik^{**}k}\mathcal{X}''(B_{k}), 
\end{align*}
where the second step follows from (\ref{decomposition}).
 Both sums on the last line include $\mathcal{X}''(B)$. 
Furthermore, in order for the inequalities to hold, 
there must be a set $B_l \in \mathcal{B}_+''$ in the left sum and a set $B_r \in \mathcal{B}_+''$ in the right sum such that 
\begin{align}
\mathcal{X}''(B_{l}) &< \mathcal{X}''(B)  < \mathcal{X}''(B_{r}). \label{strictIE}
\end{align}
These sets $B_l$ and $B_r$ are connected to $B$ via some atoms in $\mathcal{A}_{k^{*}}$ and $\mathcal{A}_{k^{**}}$, that is, $B \cup B_{l} \subseteq A_{ik^*}$ and $B \cup B_{r} \subseteq A_{ik^{**}}$ for some  $A_{ik^*} \in \mathcal{A}_{k^*}$ and $A_{ik^{**}}  \in \mathcal{A}_{k^{**}}$.

 
Now, define $\mathcal{B}_{(1)} := \{B \in \mathcal{B}_+'': \mathcal{X}''(B) = \min\{\mathcal{X}''(B^*)  : B^* \in \mathcal{B}_+''\}\}$. If some of the predictions disagree on a $B \in \mathcal{B}_{(1)}$, inequality (\ref{strictIE}) shows that there is a $B' \in \mathcal{B}_+''$ with $\mathcal{X}''(B') < \mathcal{X}''(B)$. By definition of $\mathcal{B}_{(1)}$, this is a contradiction. Assume then that the predictions agree over the atoms in $\mathcal{B}_{(1)}$. In such a case, the strict mean $\mathcal{X}''$ is equal to the unanimous prediction. This is possible only if each $B_k$ in the sum  (\ref{decomposition}) is either $B_k \in \mathcal{B}_{(1)}$ or $\P(B_k) = 0$; otherwise the corresponding $\mathcal{X}''(B_k)$ would enter the sum with positive weight and hence make it larger than the unanimous prediction. 
 Thus $\mathcal{B}_{(1)}$ is a disjoint set in a sense that there is no $A_{ij}$ such that $B \cup B' \subseteq A_{ij}$ for some $B \in \mathcal{B}_{(1)}$ and $B' \in  \mathcal{B}_+'' \setminus \mathcal{B}_{(1)}$.

Next,  define $\mathcal{B}_{(2)} := \{B \in \mathcal{B}_+'' : \mathcal{X}''(B) = \min\{\mathcal{X}''(B^*) : B^* \in \mathcal{B}_+'' \setminus \mathcal{B}_{(1)}\} \}$. Analyze $\mathcal{B}_{(2)}$ similarly to the above analysis of $\mathcal{B}_{(1)}$. In particular, if the predictions disagree on some $B \in \mathcal{B}_{(2)}$, inequality (\ref{strictIE}) leads to a contradiction because there is no $A_{ij}$ and $B' \in \mathcal{B}_+''$ such that $\mathcal{X}''(B') < \mathcal{X}''(B)$ and $B \cup B' \subseteq A_{ij}$. If, on the other hand, all predictions agree, then $\mathcal{B}_{(2)}$ is a disjoint set just like $\mathcal{B}_{(1)}$. 

Repeat this process for $\mathcal{B}_{(3)}, \mathcal{B}_{(4)}, \dots,$ and find the first set $\mathcal{B}_{(k)}$ with an atom on which some forecasters disagree. Such a set must exist because at least two forecasters are assumed to disagree with positive probability.  This then gives the final contradiction, and it can be concluded that  $\P(\mathcal{X}_{(\cdot)} \neq \mathcal{X}'') > 0$.

\qed

\subsection{Proof of Corollary \ref{asympEff}}

Given that $|\Omega| < \infty$, we can write $\P(A) = \sum_{\omega \in A} p(\omega)$, where $A \in \F$, $p(\omega) \geq 0$, and $\sum_{\omega \in \Omega} p(\omega) = 1$. In addition, given that $ \sum_{\omega \in \Omega} |\E(Y|\H_i) (\omega)| p(\omega) = \E[|\E(Y|\H_i)|]  < \infty$, the prediction $\E(Y|\H_i)$ is a.s. finite for all $i = 1, \dots, I$. Therefore  $L_X := \min(\{\E(Y|\H_i) :  \H_i \in H \text{ and } p_i >0\})$ and $M_X := \max(\{\E(Y|\H_i) :  \H_i \in H \text{ and } p_i >0\})$ are also a.s. finite.

First, define a random variable $\epsilon$ that has the same distribution as each idiosyncratic error $\epsilon_j$. Second, for a given $x$, let $\tilde{F}_x := \P(Q(x,\epsilon) \leq t)$ and $\tilde{F}_x^* := \P(Q(x,\epsilon) \geq t)$. This way each $\tilde{X}_j \sim \tilde{F} = \sum_{i=1}^I \tilde{F}_{\E(Y|\H_i)} p_i$. 
By assumption (\ref{Assump4}), for  any $x < y$  there exists some random variable $\delta >0$ such that $\tilde{F}_y(t) = \P(Q(y,\epsilon) \leq t) = \P(Q(x,\epsilon) + \delta \leq t) = \P(Q(x,\epsilon) \leq t - \delta) <  \P(Q(x,\epsilon) \leq t) = \tilde{F}_x(t)$ with $t \in \supp(\tilde{F}_x) = \{t : \P(Q(x,\epsilon) = t) > 0\}$. On the other hand, for all $t \in \supp(\tilde{F}_y)$, we have that $\tilde{F}^*_y(t) = \P(Q(y,\epsilon) \geq t) = \P(Q(x,\epsilon) \geq t- \delta)  >  \P(Q(x,\epsilon) \geq t) = \tilde{F}^*_x(t)$.
 
Next, consider the event $A_0 = \{\E(Y|\H_i) = \E(Y|\H) \text{ for some } \H \in H \text{ and all } p_i > 0\} \subseteq \Omega$. Under this event,  each $\tilde{X}_j = Q(\E(Y|\H), \epsilon_j)$ and $\mathcal{X}  \to \E(Y|\H)$. Assumption \ref{Assump5}, however, ensures that the complement $\P(A_0^c) > 0$. In other words, with positive probability there exist some  $i \neq i'$ such that $\E(Y|\H_i) \neq \E(Y|\H_{i'})$ and $p_i, p_{i'} > 0$.  Thus, for each $\omega \in A_0^c$, $L_X < \E(Y|\H_i)$ and $\E(Y|\H_{i'}) < M_X$ for some $i,i'$. Consequently,
$\tilde{F}(t) = \sum_{i=1}^I \tilde{F}_{\E(Y|\H_i)}(t) p_i <  \sum_{i=1}^I \tilde{F}_{L_X}(t) p_i  = \tilde{F}_{L_X}(t)$ 
 for all $t \in \supp(\tilde{F}_{L_X})$, and
$\tilde{F}^*_{M_X}(t) =  \sum_{i=1}^I \tilde{F}^*_{M_X}(t) p_i > \sum_{i=1}^I \tilde{F}^*_{\E(Y|\H_i)}(t) p_i = \tilde{F}^*(t)$ 
  for all $t \in \supp(\tilde{F}_{M_X})$.
  Now, by Assumption \ref{Assump3}, $L_X = T(\tilde{F}_{L_X}) < T(\tilde{F}) < T(\tilde{F}_{M_X}) = M_X$. Consequently, $\mathcal{X} \to T(\tilde{F})  \in (L_X, M_X)$ under $A_0^c$. Putting this together shows that $\mathcal{X}$ is asymptotically a strict mean of the calibrated predictions $\E(Y|\H_i)$. Given that all $|\F_j| < \infty$, by Theorem \ref{centralTendency}, $\mathcal{X}$ is not asymptotically efficient. 
\qed


\bibliographystyle{apalike}
\bibliography{biblio}		

\end{document}